# A Redox-based Ion-Gating Reservoir, Utilizing Double Reservoir States in Drain and Gate Nonlinear Responses

*Tomoki Wada,*[#,†,‡] *Daiki Nishioka,*[#,†,‡] *Wataru Namiki*[†], *Takashi Tsuchiya,*[*,†,‡] *Tohru Higuchi*[‡] *and Kazuya Terabe*[†]

[†]International Center for Materials Nanoarchitectonics (WPI-MANA), National Institute for Materials Science (NIMS), 1-1 Namiki, Tsukuba, Ibaraki 305-0044, Japan

[‡]Department of Applied Physics, Faculty of Science, Tokyo University of Science, 6-3-1 Niijuku, Katsushika, Tokyo 125-8585, Japan

[#]Equally contributed
*Email: TSUCHIYA.Takashi@nims.go.jp

Abstract: We have demonstrated physical reservoir computing with a redox-based ion-gating reservoir (redox-IGR) comprising $Li_xWO_3$ thin film and lithium ion conducting glass ceramic (LICGC). The subject redox-IGR successfully solved a second-order nonlinear dynamic equation by utilizing voltage pulse driven ion-gating in a $Li_xWO_3$ channel to enable reservoir computing. Under the normal conditions, in which only the drain current ($I_D$) is used for the reservoir states, the lowest prediction error is $7.39\times10^{-4}$. Performance was enhanced by the addition of $I_G$ to the reservoir states, resulting in a significant lowering of the prediction error to $5.06\times10^{-4}$, which is noticeably lower than other types of physical reservoirs reported to date. A second-order nonlinear autoregressive moving average (NARMA2) task, a typical benchmark of reservoir computing, was also performed with the IGR and good performance was achieved, with an NMSE of 0.163. A short-term memory task was performed to investigate an enhancement mechanism resulting from the $I_G$ addition. An increase in memory capacity, from 1.87 without $I_G$ to 2.73 with $I_G$, was observed in the forgetting curves, indicating that enhancement of both high dimensionality and memory capacity are attributed to the origin of the performance improvement.

Recently, the research on and development of physical reservoir computing has seen increased activity due to the tremendous potential for it to significantly reducing computation resources compared to conventional machine learning approaches, which are based solely on semiconductor integrated circuits.[1,2] Various materials and devices, including soft bodies, optical devices, analogue circuits, spin torque oscillators, memristors, nanowire networks, and ion-gating transistors, have been reported to function as physical reservoirs that require nonlinearity, high dimensionality, and short-term memory,[2-25] while the demonstrated computing performance has been far from satisfactory to date. One common characteristic of a physical reservoir that is proving to be extremely difficult to achieve is the securing of high dimensionality, which is in essence the obtaining of a sufficient number of reservoir states from the output of a physical reservoir. This is because the outputs of physical reservoirs are measured as small numbers of time-series responses with a limited number of detecting probes (e.g., electrodes, sensors), which are attached to or arranged in some manner with the reservoir under serious geometrical constraints. This is in direct contrast to fully-simulated reservoirs, in which unrestricted access to the reservoir states of nodes is enabled. Virtual node methods are useful in compensating for said lack of high dimensionality, and are thus widely used.[7-10] In such method, post processing allows for a lot of virtual nodes to be obtained from given time-series data. However, there is a known trade-off relationship between increasing the number of virtual nodes and the diversity of each virtual node.[26] It is thus not straightforward to secure a sufficient number of diverse virtual nodes from a limited number of time-series response, which makes physical reservoirs impracticable. Therefore, to achieve practical use, it is necessary to explore physical reservoirs that have diverse outputs.

Here, we report a redox-ion-gating reservoir comprised of all-solid-state redox transistors,[27-40] which can derive double reservoir states from drain and gate current response, based on ion-insertion and desertion (redox) through a solid electrolyte,[41-45] at a Li$^+$-electron mixed conductor, Li$_x$WO$_3$. Using sequential gate voltage pulse trains, a drain current (electronic current) flows through a Li$_x$WO$_3$ thin film channel, where it is modified by a redox reaction with a Li$^+$ conducting glass ceramic (LICGC) substrate through the modulation of the conducting electron density, so as to generate a nonlinear time-series response in the drain current. Simultaneously, a relatively large gate current for the redox process (lithium ion current) can provide another time-series response. In the normal measurement configuration of transistor devices, two responses (drain and gate) with different characteristics are easily obtained. This increases the number of virtual nodes by overcoming the said trade-off relationship. By employing a redox-ion-gating reservoir, second-order nonlinear dynamical tasks and a second-order nonlinear autoregressive moving average (NARMA2) were successfully solved, with normalized mean square errors (NMSEs) of $5.06 \times 10^{-4}$ and 0.163, respectively. Said ion-gating reservoir structure, with inorganic materials, is useful as a building block for semiconductor integrated circuits. Therefore, it is shown that the approach described herein can contribute to the physical implementation of physical reservoirs in practical devices that require compatibility together with high computational performance and high density integration.

RESULTS AND DISCUSSION

**General concept of the redox-based, ion-gating reservoir.** Figure 1 (a) is a schematic diagram of the redox-based, ion-gating reservoir (redox-IGR) in the study. Using RF sputtering, we deposited on a 0.15mm-thick LICGC substrate a $LiCoO_2$ thin film (200 nm), with huge storage capacity Li ion/Pt thin film (50 nm) as a gate electrode, Pt thin film (50 nm) as drain and source electrodes, and $WO_3$ thin film (100 nm) as a channel, respectively. In order to insert Li ions into the $WO_3$ channel, a constant voltage of 2.5 V was applied between the gate and source electrodes. The $I_D$ (drain current)-$V_G$ (gate voltage) and $I_G$ (gate current)-$V_G$ characteristics of the redox-based IGR are shown in Fig. 1 (b) and (c). The gate voltage is swept from 0.5 V to 1.5 V and then back to 0.5 V at various sweep rates, ranging from 5 mV/s (slow) to 250 mV/s (fast). $I_D$ is normalized at the initial value for comparison to those measured under different sweeping rate conditions. The $I_D$ is modulated by the application of $V_G$ because the conducting electron is doped (or removed) in the $Li_xWO_3$ by the redox reaction ($Li^+$ insertion or desertion) (Eq. 1).

$$Li_xWO_3 + yLi^+ + ye^- \leftrightarrow Li_{x+y}WO_3 \qquad (1)$$
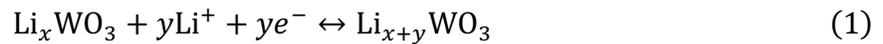

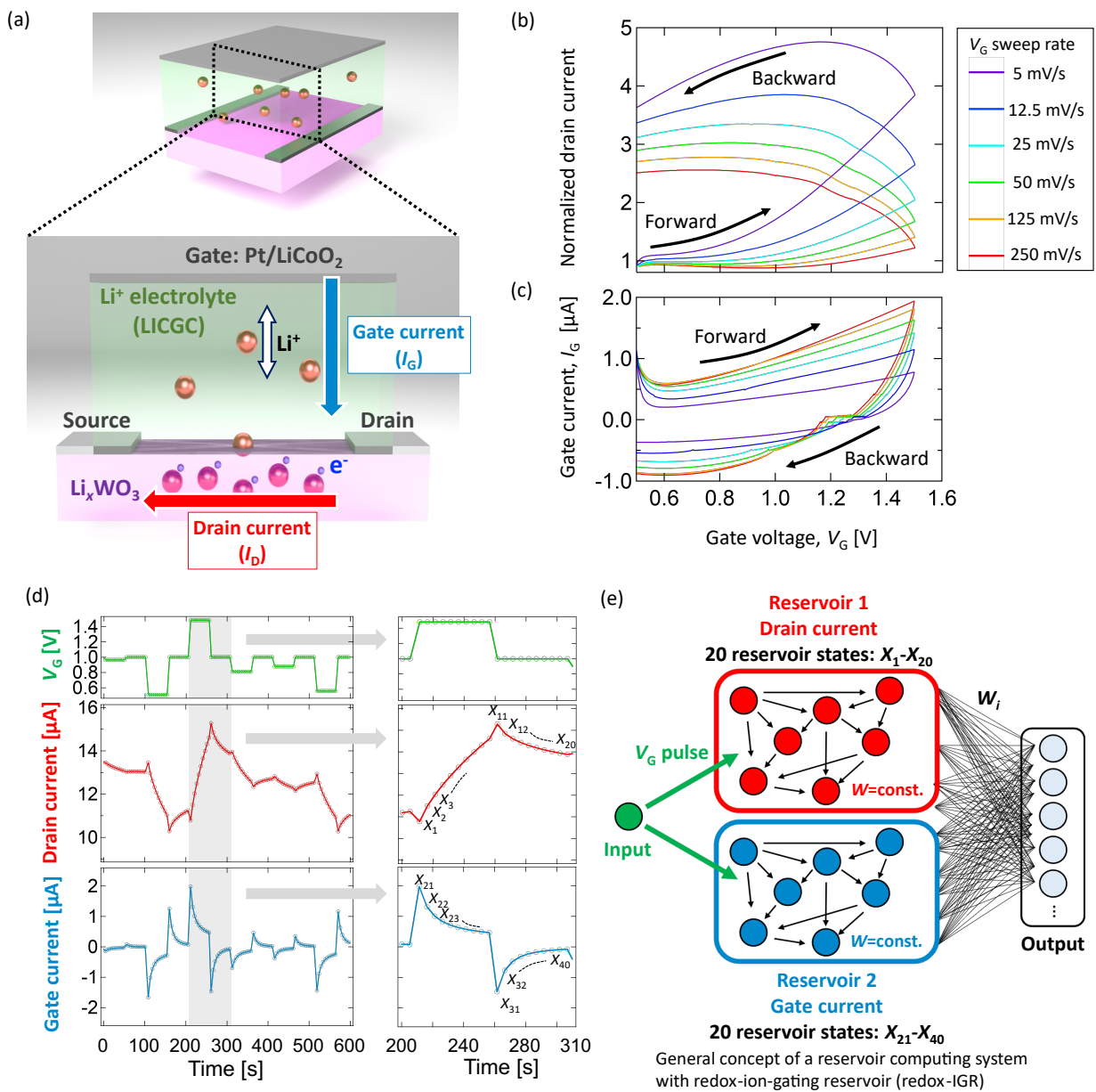

**Figure 1.** (**a**) Schematic image of a $Li_xWO_3$-based redox-ion-gating reservoir. (**b**) Normalized drain current, and (**c**) gate current measured during $V_G$ sweeping from 0.5 to 1.5 V. (**d**) Gate voltage pulse stream, drain current response, and gate current response during operation of the redox-ion-gating reservoir. 40 reservoir states $X_i$ ($i = 1, …, 40$) are obtained as shown in the right panels. (**e**) General concept of a reservoir computing system with redox-ion-gating reservoir. $W_i$ denotes the read-out weight.

The $I_D$-$V_G$ curves are accompanied by smaller hysteresis as the sweep rate raises. In the subject redox-IGR transistor, Li$^+$ transport in the Li$_x$WO$_3$ channel is much slower than in the electrolyte: a rate limiting step of the overall Li$^+$ transport is Li$^+$ transport in the Li$_x$WO$_3$ channel. As the sweep rate rises, the delay of Li$^+$ transport in the Li$_x$WO$_3$ to gate voltage sweep becomes large, resulting in large hystereses in the $I_D$-$V_G$ curves. Therefore, the $I_D$-$V_G$ curve shows different hysteresis characteristics depending on the $V_G$ sweep rate, which is the origin of the short-term memory of the redox-IGR.

To perform time-series tasks using the transistor, $I_D$ response to gate voltage pulse is useful for mapping input signals to higher dimensional feature space. $V_G$ pulse streams can be used to deal with sequential time-series signals. The upper and middle panels of Fig. 1 (d) show an example of $I_D$ response (middle) with respect to $V_G$ pulse streams (upper), which are input signals to the transistor. When one $V_G$ pulse (corresponding to one point in a time-series data-set) is input, 20 reservoir states $X_i$ ($i = 1, \ldots, 20$) can be obtained from the $I_D$ response by the virtual node method, as shown in the right panels of Fig. 1 (d). Conventional ion-gating reservoirs (IGR) use only the $I_D$,[21] but the subject redox-IGR can use $I_G$ as a reservoir state. This is due to the significant $I_G$ response, which is completely different from the $I_D$ response, as evidenced in Fig. 1 (d). Furthermore, the $I_G$ is much larger than the one found in a electric double layer-IGR.[21] Therefore, another 20 reservoir states $X_i$ ($i = 21, \ldots, 40$) can be obtained from the $I_G$ response, as shown in the lower and right panels of Fig. 1 (d). The doubled reservoir states can be utilized to perform reservoir computing with enhanced high dimensionality, as schematically shown in Fig. 1 (e). Details of origin in the difference of $I_D$ and $I_G$ responses will be discussed later. In the following sections, physical reservoir computing performance with the responses with specific tasks will be shown.

**Solving a second-order nonlinear dynamic equation.** Reservoir computing is advantageous for time-series data analysis due to the nonlinearity, short-term memory, and high dimensionality of the reservoir for input signals. Therefore, we evaluated the computational performance of the subject redox-IGR in time series data analysis by solving a second-order nonlinear equation task. The general concept of a process flow diagram for the second-order nonlinear equation task solved is shown in Fig. 2 (a).[7,8] The target time series $y_t(k)$ for this task is generated by the second-order nonlinear dynamic equation shown in Eq.2.

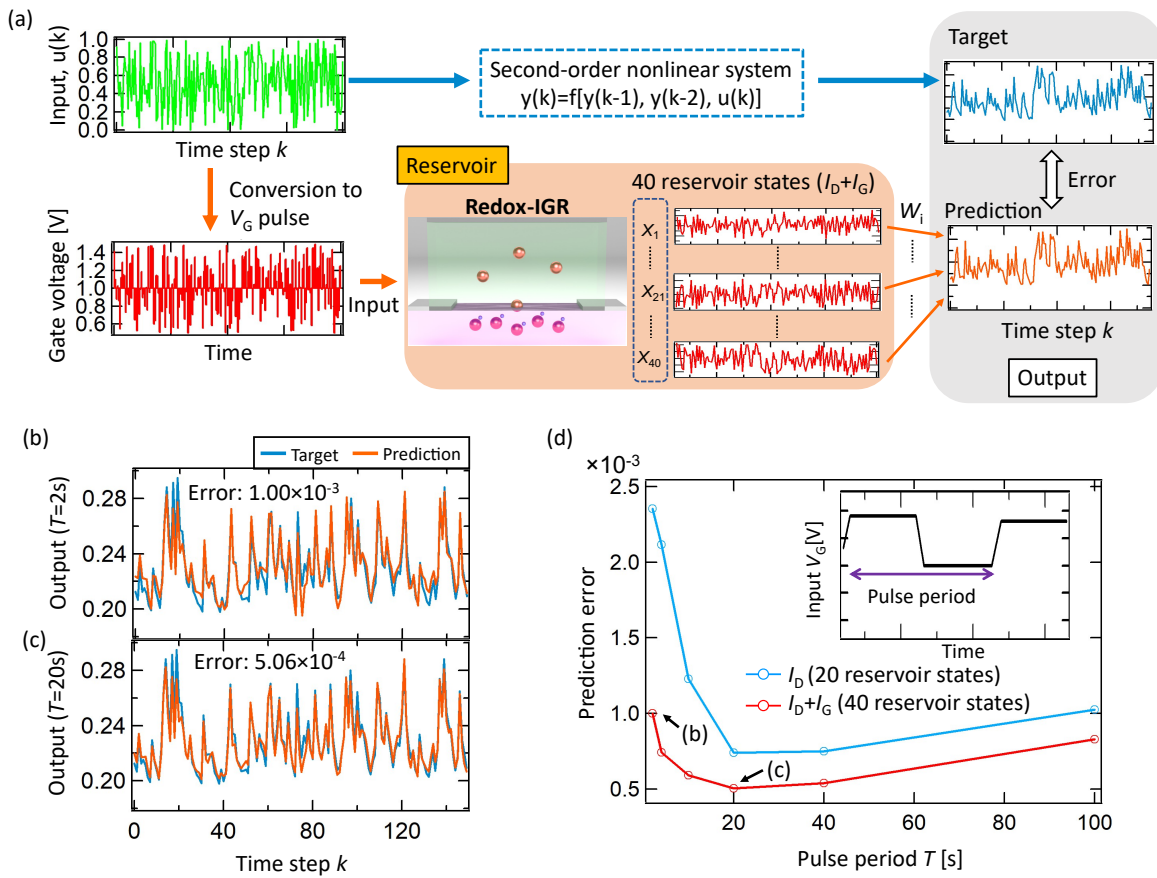

**Figure 2.** (**a**) Process flow diagram of a second-order nonlinear equation task, showing the target (blue line) and prediction (orange line) waveforms at (**b**) $T = 2$ s and (**c**) $T = 20$ s. (**d**) Relationship between prediction error and pulse interval under conditions with only $I_D$ (blue line) and with $I_D+I_G$ (red line). The pulse period is defined as shown in Fig. 2 (d) inset.

$$y_t(k) = f[y_t(k-1), y_t(k-2), u(k)]$$

$$= 0.4y_t(k-1) + 0.4y_t(k-1)y_t(k-2) + 0.6u^3(k) + 0.1 \quad (2)$$

, where $u(k)$ and $k$ are a random input ranging from 0 to 0.5 and the discrete time, respectively. Equation 2 contains a second-order nonlinearity and a two-step forward term which, in order to solve the equation, which elements are require to be expressed by the reservoir as linearly separable.[7]

A random input $u(k)$ was linearly converted to a voltage pulse stream, with a pulse period of $T$ (2 to 100 s) and a duty rate of 50%, which was input to the subject redox-IGR transistor under a constant $V_D$ of 0.1 V. The pulse intensity $V_G(k)$ ranged from 0.5 V to 1.5V ($V_G(k) = 2u(k) + 0.5$ V) and a constant $V_G$ of 1 V was applied during the pulse interval. The gate voltage pulse stream, drain current response, and gate current response are shown in Fig. 1 (d). As already discussed, a total of 40 reservoir states $X_i$ were obtained from both the $I_D$ ($i = 1, \ldots, 20$) and the $I_G$ ($i = 21,\ldots,40$) responses by the virtual node method.

By combining physical and virtual nodes with different characteristics, the unique electrical behavior of the subject redox-IGR, caused by redox reactions (electronic current through channel and ion currents through electrolyte, which are associated with $Li^+$ transport), can be extracted as reservoir states that are valid for reservoir computing and can further be mapped to a high-dimensional feature space.[21] As a result, the reservoir output $y(k)$ is obtained by following equation,

$$y(k) = \sum_{i=1}^{N} w_i X_i(k) + b \qquad (3)$$

, where $N$, $w_i$ and $b$ are the size of the reservoir (= 40), read-out weights and bias, respectively. See the Methods section for additional details on the learning algorithm.

In the test phase, in order to evaluate the generalization performance of the subject redox-IGR, we checked whether the reservoir output (Eq. 3) with fixed $w_i$ matched the learned equation (Eq. 2) for inputs different from those in the training phase. The prediction error defined below was used to evaluate the computational performance of the subject redox-IGR in this task.

$$\text{Prediction error} = \frac{\sum_{k=1}\{y_t(k) - y_p(k)\}^2}{\sum_{k=1} y_t(k)^2} \qquad (4)$$

Figures 2 (b) and (c) show the target and predicted waveforms when the subject redox-IGR transistor was operated at different pulse periods $T$ of 2 s and 20 s in the test phase. It is particularly noteworthy that the target and predicted waveforms are in excellent agreement for $T = 20$ s, as shown in Fig. 2 (c). That is, Eq. 2 was successfully solved by our redox-IGR, with a prediction error of $5.06 \times 10^{-4}$, which is sufficiently low in comparison to other physical reservoirs reported to date ($1.31 \sim 3.13 \times 10^{-3}$).[7,8] On the other hand, the prediction error worsened when the subject IGR was operated with $T$ of 2 s, as shown in Fig. 2 (b). This is because the relaxation process in the subject redox-IGR is correlated with the sweep rate of the gate input, as detailed in Fig. 1. To evaluate the correlation between the operating conditions of the subject redox-IGR and its computational performance as a reservoir, we investigated the relationship between $T$ and the prediction error, as shown in Fig. 2 (d). The blue line in the figure shows the results when only the $I_D$ is used for the reservoir states (i.e., $X_i$, $i = 1, ..., 20$), and the red line shows the results when both the $I_G$ and the $I_D$ are used for the reservoir states (i.e., $X_i$, $i = 1, ..., 40$). It is found that the prediction performance is best at $T = 20$ s, regardless of the presence or absence of a gate current. This is because, when the pulse period is short, the redox reaction shown in Eq. 1 proceeds with too great a delay, and the ion current associated with ion transport in the electrolyte dominates. Therefore, the short-term memory characteristics and nonlinearity due to the resistance modulation of $Li_xWO_3$ are lost, and the only current response obtained is a simple one similar to the relaxation process of a Resistor-Capacitor parallel circuit. In addition, if the pulse period is too long, the interaction between the virtual nodes is suppressed, which results in poor computational performance.[10]

The use of $I_G$ in addition to $I_D$ not only lowers the said error, but also moderates the dependence of the computational performance on the input conditions. This is because, in addition to increased expressive power due to the increased reservoir size, the $I_D$-derived $X$ and $I_G$-derived $X$ utilize complementary features that are necessary for task execution, which is an important feature of the subject redox-IGR, which utilizes different physical nodes as computational resources. This feature of the subject redox-IGR, which provides good computational performance regardless of slight changes in its operating conditions, is also an extremely significant practical advantage for reservoir computing implementation.

**Evaluation of prediction performance for a NARMA 2 task.** To further evaluate the performance of time series prediction using the subject redox-IGR, we have performed a NARMA2-task, which is more difficult than the second-order nonlinear equation task performed in the previous section, as well as being a typical benchmark task for both full-simulation reservoir computing and physical reservoir computing[4-6,22-25] The time series prediction generated by the NARMA2 model, with specific parameters defined by equation (5), is a popular benchmark for the development of physical reservoirs.[4-6,22,23]

$$y_t(k+1) = 0.4y_t(k) + 0.4y_t(k)y_t(k-1) + 0.6u^3(k) + 0.1 \tag{5}$$

Figures 3 (a) and (b) show the results of waveform prediction with both drain and gate currents at $T = 2$ s and 40 s, respectively. The error value, a normalized mean square error (NMSE), between the target waveform (blue line) and the predicted waveform (orange line) is defined by Eq. (6)

$$\text{NMSE} = \frac{1}{L} \times \frac{\sum_{k=1}^{L}(y_t(k) - y(k))^2}{\sigma^2(y_t(k))} \tag{6}$$

where $L$ (= 150) is a data length. The computational performance of the subject redox-IGR in performing the NARMA2 task is enhanced by the higher dimension, combined with $I_G$. As seen in the comparison of the two conditions ($T = 2$ and 40 s) with $I_G$ shown in Figs. 3 (a) and (b), the minimum value of NMSE is 0.163 at $T = 40$ s, whereas the maximum NMSE is 0.321 at $T = 2$ s. Figure 3 (c) shows the relationship between NMSE and pulse interval under conditions with only $I_D$ (blue line) and with $I_D+I_G$ (red line). The result, in which utilizing $I_G$ in addition to $I_D$ gives better performance over the whole pulse period range, is quite similar to the case for the second-order nonlinear equation task, which gives support to our conviction that the present approach is sufficiently versatile to achieve an information processing ability in the subject redox-IGR.

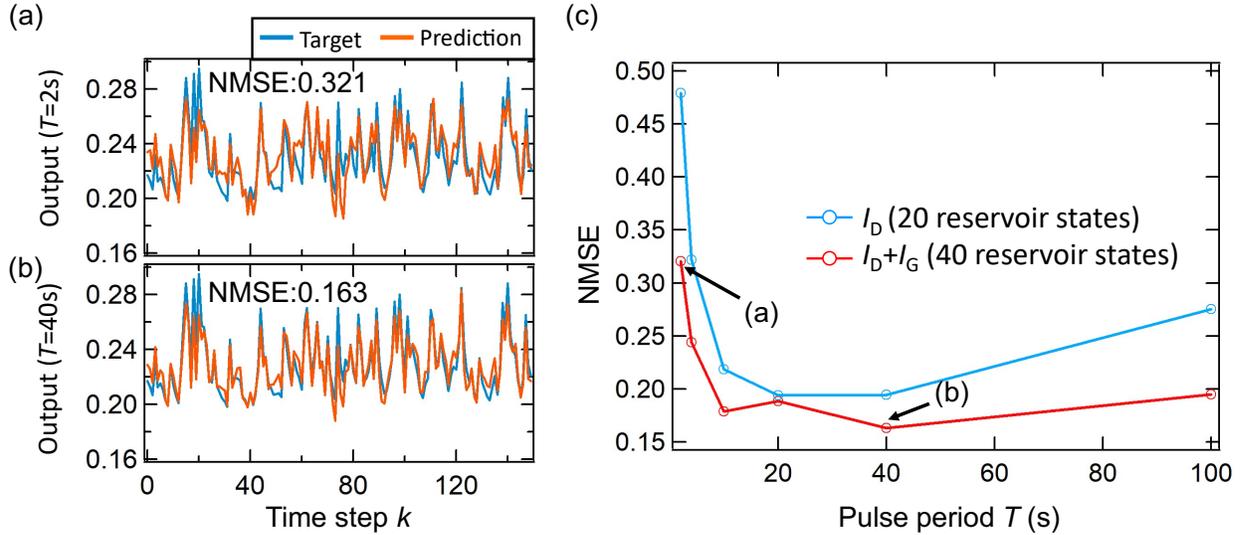

**Figure 3.** The target (blue line) and prediction (orange line) waveforms at **(a)** $T$=2s and **(b)** $T$=40s. **(c)** Relationship between NMSE and pulse interval. The blue line and the red line represent 20 and 40 reservoir states, respectively.

**Evaluation of Memory Capacity.** In order to further investigate the underlying mechanism of the enhancement effect elicited by the addition of $I_G$ to the reservoir states, we performed a short-term memory task, which task measures the ability of our redox-IGR to reconstruct past time series data input to the redox-IGR. Here, as in the time series analysis task described in Figs. 2 and 3, a voltage-transformed random input $u(k)$ is applied to the subject redox-IGR and the input $u(k-\tau)$ before the delay time $\tau$ is reconstructed by a linear combination of reservoir states and weights obtained from the current response of the subject redox-IGR (Eq. 3). The agreement between the target waveform $u(k-\tau)$ and the reconstructed waveform $y(k)$ by the reservoir was evaluated using the following coefficient of determination $r^2$,

$$r^2(\tau) = \frac{\text{Cov}^2\big(u(k-\tau), y(k)\big)}{\text{Var}\big(u(k)\big) \times \text{Var}\big(y(k)\big)} \qquad (7)$$

,where Cov() and Var() are the covariance and variance, respectively. Figure 4 (a) shows the forgetting curve (determination coefficient vs. delay) of the subject redox-IGR using $I_D$, $I_G$ and $I_D+I_G$ at $T = 20$ s.[5] The determination coefficient $r^2$ (i.e., the ability for reconstruction) decreases as the delay increases. That is a universal feature of short-term memory. Memory capacities (MC) for the three conditions are calculated to be 1.87 for $I_D$, 2.00 for $I_G$ and 2.73 for $I_D+I_G$, respectively, by integration of the curves in Fig. 4 (a) as follows,

$$\mathrm{MC} = \sum_{\tau=1}^{\infty} r^2(\tau) \tag{8}$$

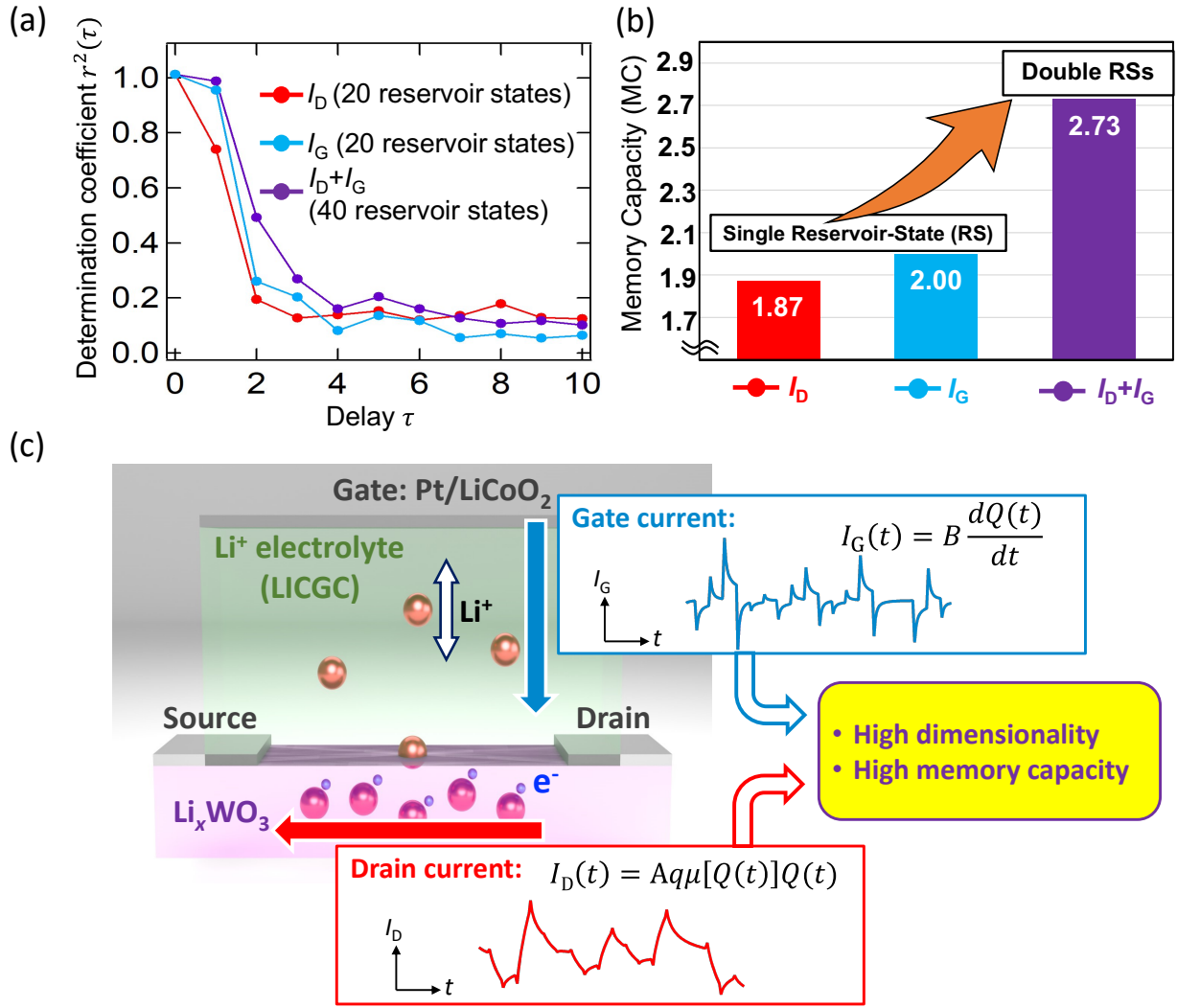

**Figure 4.** (a) Forgetting curves of the subject IGR using $I_D$, $I_G$ and $I_D+I_G$, indicated by red, blue and purple lines, respectively. (b) Memory capacity (MC) for each condition. (c) Comparison with drain and gate currents, expressed as a function of $Q(t)$.

By comparing with MCs under the three conditions, it is revealed that the double reservoir states combining $I_G$ and $I_D$ enhance not only high dimensionality but also the MC of the reservoir.

Given $Q(t)$ as total charges (Li$^+$) stored in Li$_x$WO$_3$, $I_G$ is expressed by $\frac{dQ(t)}{dt}$. On the other hand, $I_D$ is expressed by $Aq\mu[Q(t)]Q(t)$, in which A, $q$, and $\mu[Q(t)]$ are a constant, elementary charge, and electron mobility, respectively. This comparison is illustrated in Fig. 4 (c). Whereas both the $I_D$ and $I_G$ include $Q(t)$ as a component, they differently contribute to the current as is clearly found in the expressions. Furthermore, electron mobility $\mu[Q(t)]$ can be dependent on $Q(t)$ if one considers an interference of electronic conduction by concentrated Li$^+$ in a WO$_3$ matrix. Therefore, $I_D$ and $I_G$ can provide different reservoir states, although the two are coupled in the carrier doping process that is shown in Eq. 1. The combination of the provided reservoir states generates said high dimensionality and MC, which combination provides improved information processing performance.

CONCLUSIONS

Physical reservoir computing, with a redox-IGR comprised of Li$_x$WO$_3$ thin film and LICGC, has been demonstrated. The subject redox-IGR successfully solved a second-order nonlinear dynamic equation, with a lowest prediction error of 7.39×10$^{-4}$, under a normal condition where only $I_D$ is used for reservoir states. Performance was enhanced by the addition of $I_G$ to the reservoir states, resulting in a significant lowering of the prediction error to 5.06×10$^{-4}$, which is noticeably lower than other types of physical reservoirs reported to date. NARMA2, a typical reservoir computing benchmark, was also performed with the subject redox-IGR. Better performance was achieved, with an NMSE of 0.163, by the addition of $I_G$ to the reservoir states, which reveals that $I_G$ is a useful source for obtaining better reservoir properties. A short-term memory task was performed to investigate enhancement mechanism resulting from the addition of $I_G$. The forgetting curves of the subject redox-IGR show that MC was enhanced from 1.87 with $I_D$ to 2.73 with $I_D+I_G$. The enhancement of both high dimensionality and MC resulting from the addition of $I_G$ to the reservoir states is attributed to the origin of the performance improvement.

Physical reservoir computing is, from a certain viewpoint, an attempt to utilize as many inherent properties of a material/device as is possible so as to achieve efficient information processing. For a transistor, $I_G$ is usually regarded as of no use, nevertheless it includes certain internal and temporal information about the transistor. The present technique is useful in harnessing such internal information in a device so as to realize the efficient mapping of input to higher dimensional feature space. This approach can be applied to a wide range of multi-component physical reservoir systems.

METHODS

**Fabrication of a Li$_x$WO$_3$-based redox transistor.** The Li$_x$WO$_3$-based redox transistor, schematically shown in Fig 1(a), was fabricated on a 0.15 mm-thick LICGC substrate. A 100 nm-thick WO$_3$ thin film was deposited by the RF sputtering method, using a sintered stoichiometric WO$_3$ target with 99.9 % purity, with a supply of pure Ar and O$_2$ gases at fixed flow rates of 10 and 0.6 sccm, respectively. Then, a 200 nm-thick LiCoO$_2$ thin film was deposited as a gate electrode on the opposite side of the LICGC substrate, against the WO$_3$ side, with a supply of pure Ar and O$_2$ gases at fixed flow rates of 9 and 3 sccm, respectively. The drain and source electrodes and the current collector on the gate electrode were made of 50 nm-thick Pt films. All film was deposited by the RF sputtering method at room temperature. Prior to measurements being made, a constant voltage of 2.5 V was applied between the gate and source electrodes so as to insert Li ions into the WO$_3$ channel.

**Measurement of $I_D$ and $I_G$ responses.** All electrical measurements of the subject redox-IGR were carried out at room temperature in a vacuum chamber and carried out using the source measure unit (SMU) of a semiconductor parameter analyzer (4200A-SCS, Keithley). A random input $u(k)$ was linearly converted to the voltage pulse streams, with a pulse period of $T$ (2 to 100 s) and a duty rate of 50%, which was input to the subject redox-IGR transistor under constant $V_D$ of 0.1 V. The pulse intensity $V_G(k)$ ranged from 0.5 V to 1.5V ($V_G(k) = 2u(k) + 0.5$ V) and constant $V_G$ of 1 V was applied during pulse intervals. The $I_D$ and $I_G$ responses of the subject redox-IGR were monitored, and 20 virtual nodes were extracted from each response. Thus, 40 reservoir states were obtained from input $u(k)$ by the subject redox-IGR. Said reservoir states were normalized from 0 to 1 for calculation, as shown in Eq. (3).

**Ridge regression for time-series data analysis tasks.** In the time series data analysis tasks, such as the solving of the second-order nonlinear dynamic task and the NARMA2 task shown in Fig. 2 and Fig. 3, the readout network of the subject redox-IGR was trained by ridge regression. Here we describe the algorithm used for said ridge regression. The reservoir output $y(k)$ shown in Eq. 3 can also be defined as follows;

$$y(k) = \boldsymbol{w} \cdot \boldsymbol{x}(k) \tag{9}$$

, where $\mathbf{w} = (b, w_1, ..., w_N)$ and $\mathbf{x}(k) = (1, X_1(k), ..., X_N(k))^T$ are the weight vector and the reservoir state vector with a reservoir size of $N$, respectively. The cost function $J(\mathbf{W})$ in ridge regression is defined as follows

$$J(\mathbf{w}) = \frac{1}{2}\sum_{k=1}^{L}(y_t(k) - y(k))^2 + \frac{\beta}{2}\sum_{i=0}^{N} w_i^2 \quad (10)$$

, where $L$, $\lambda$ and $y_t(k)$ are the data length in the training phase, the ridge parameter and the target output generated by Eq. 2 or Eq. 5, respectively. The data length and ridge parameter were $L = 150$ and $\beta \leq 2 \times 10^{-3}$ for all the tasks detailed in Fig. 2 and Fig. 3. The trained weights $\widehat{\mathbf{w}}$ that minimize cost function $J(\mathbf{w})$ is given by following equation.

$$\widehat{\mathbf{w}} = \mathbf{YX}^T(\mathbf{XX}^T + \lambda \mathbf{I})^{-1} \quad (11)$$

, where $\mathbf{Y} = (y_t(1), y_t(2), ..., y_t(L))$, $\mathbf{X} = (\mathbf{x}(1), \mathbf{x}(2), ..., \mathbf{x}(L))$ and $\mathbf{I}(\subseteq \mathbb{R}^{(N+1)\times(N+1)})$ are the target output vector, the reservoir state matrix and the identify matrix, respectively.

ASSOCIATED CONTENT

AUTHOR INFORMATION

**Corresponding Author**

*E-mail: TSUCHIYA.Takashi@nims.go.jp

ACKNOWLEDGMENT

This work was in part supported by Japan Society for the Promotion of Science (JSPS) KAKENHI Grant Number JP22H04625 (Grant-in-Aid for Scientific Research on Innovative Areas "Interface Ionics"), and JP21J21982 (Grant-in-Aid for JSPS Fellows). A part of this work was supported by the Yazaki Memorial Foundation for Science and Technology and Kurata Grants from The Hitachi Global Foundation.